\documentclass[conference]{IEEEtran} 
\usepackage{epsfig,graphicx,color}

\begin{document}

  \title{A File System Abstraction for Sense and Respond Systems}

  % Author names and affiliations.  Use a multiple column layout
  % for up to three different affiliations.
%  \author{\authorblockN{the first}
%    \authorblockA{School of Electrical and\\Computer Engineering\\
%      Indiana University\\
%      Atlanta, Georgia 30332--0250\\
%      Email: mshell@ece.gatech.edu}
%    \and
%    \authorblockN{the second}
%    \authorblockA{School of Computer Science\\
%      Binghamton University\\
%      Email: homer@thesimpsons.com}
%  }

\author {Sameer Tilak$^{1}$, Bhanu Pisupati$^{2}$, Kenneth Chiu$^{1}$, Geoffrey Brown$^{2}$, Nael Abu-Ghazaleh$^{1}$ \\
{{$^{1}$Computer Science Department, State University of New York (SUNY) at Binghamton}} \\
 $^{2}$Computer Science Department, Indiana University \\}

% Author names and affiliations.  Use a multiple column layout
% for up to three different affiliations.
% \author{\authorblockN{Michael Shell}
% \authorblockA{School of Electrical and\\Computer Engineering\\
% Georgia Institute of Technology\\
% Atlanta, Georgia 30332--0250\\
% Email: mshell@ece.gatech.edu}
% \and
% \authorblockN{Homer Simpson}
% \authorblockA{Twentieth Century Fox\\
% Springfield, USA\\
% Email: homer@thesimpsons.com}
% \and
% \authorblockN{James Kirk\\ and Montgomery Scott}
% \authorblockA{Starfleet Academy\\
% San Francisco, California 96678-2391\\
% Telephone: (800) 555--1212\\
% Fax: (888) 555--1212}}

  \maketitle
  
  \begin{abstract}
    The heterogeneity and resource constraints of sense-and-respond
    systems pose significant challenges to system and application
    development.  In this paper, we present a flexible, intuitive file
    system abstraction for organizing and managing sense-and-respond
    systems based on the Plan 9 design principles.  A key feature of
    this abstraction is the ability to support multiple views of the
    system via filesystem namespaces.  Constructed logical views
    present an application-specific representation of the network,
    thus enabling high-level programming of the network.  Concurrently,
    structural views of the network enable resource-efficient planning
    and execution of tasks.  We present and motivate the design using
    several examples, outline research challenges and our research
    plan to address them, and describe the current state of implementation. 
  \end{abstract}

%   \section{Introduction}  
  \section{Introduction}

The heterogeneity and resource constraints of typical sense-and-respond
(S\&R) systems pose daunting challenges to system and application
development.
These challenges are further exacerbated by the lack of simple
abstractions for the use and development of these systems.
In this paper,
we show how the principles of Plan 9~\cite{pike95plan} can be applied to S\&R systems,
resulting in flexible, intuitive systems supporting multiple logical views.
Applications can then use the view with the most appropriate
organization and abstraction.

Sense-and-respond systems typically comprise a diverse set of hardware
and software elements.
Hardware elements include a wide variety of different sensor and actuator
types of differing origins,
ranging from COTS to highly-specialized, one-of-a-kind parts.
Software elements draw from numerous domains,
including the natural sciences, artificial intelligence, 
sensor networks, and embedded systems.
Further increasing the diversity is the various ways
in which the software and hardware elements may interact,
such as event-driven, polled data, or data streams.
This heterogeneity greatly complicates the development of
of reliable, effective S\&R systems.

\begin{figure}
  \centering
\epsfig{file=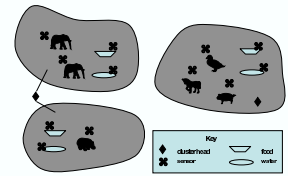,width=0.50\linewidth,silent=}
  \caption{An example wireless sensor network in a zoo.
  Sensors track animal locations and resources such as food and water.
  The network is divided into two clusters,
  each consisting of a cluster head.}
  \label{zoo_diag}
\end{figure}

A crucial component of many S\&R systems is wireless sensor and actuator
networks.
These networks promise to revolutionize sensing across a wide range of
civil, scientific, military, and industrial applications.
For example,
thousands of sensors can be deployed across
the landscape to monitor for chemical and biological threats,
or to monitor for interesting ecological events
in migration patterns~\cite{juang-02},
or to track a smoldering forest fire for conditions that might
lead to a an outbreak.
Responses may range from alerts to the use of actuators to
mitigate the damage.

The inherent resource constraints of WSNs pose significant challenges
to this vision, however.
Wireless sensors are typically limited in power, weight, and size;
and communication is often unreliable.
These constraints further exacerbate the problems created by the
heterogeneity of S\&R systems.

Successfully addressing these multi-dimensional challenges relies crucially
on developing an effective abstraction for sensor networks.
A simple and well understood abstraction 
can significantly ease both system development and application
development.
Many sensor network are deployed by scientists and
researchers whose domain of expertise is not computer science.
Motivated by this need,
we propose a simple yet powerful 
filesystem-based abstraction of sensor networks based
on Plan 9,
which espoused that
the file system metaphor
(as seen, for example, in the \verb+/proc+ file system)
can be adopted for almost all aspects of system design and development.
Not only can files be used to store a named sequence of bytes,
but also to replace many aspects of communication and control that are
typically performed using system calls.
A key feature of our proposed
solution is the ability of the application to define namespaces to 
organize the sensor network in an application specific manner.
% In fact,
% multiple concurrent perspectives of the network are possible.
Another advantage of a file system abstraction is that we can now
exploit, perhaps with some adaptation,
much of the work in distributed file systems, such as Coda~\cite{kistler-92}.
%Throughout 
%this paper, with the help of concrete examples, we demonstrate how 
%a range of canonical applications can be built around the concept of namespace
%to provide richer yet resource efficient interfaces to sensor network users.

Another commonly proposed abstraction of WSNs is that of a database~\cite{yao-02}.
Typically,
these databases present application-level information,
decoupling it from the resources.
Query processing then lacks application knowledge,
precluding the application of the end-to-end principle
and complicating efficient implementations.
This lack of low-level information in the abstraction also prevents the
provisioning of infrastructure services.
By providing logical and structural namespaces,
the file system gives the application complete control on task planning and
execution if it so desires.
%While such an abstraction may be appropriate for some applications,
%relational databases are not well-suited to representing
%irregular hierarchical structures such as might be useful for WSNs.

We model sensor networks as a set of clusters,
each with a cluster head.
Cluster membership is normally determined geographically.
Our model is intended merely to provide a concrete basis
for demonstrating the utility of our file system abstraction,
and not as an end unto itself.
With this abstraction,
an application might access sensor data geographically
by reading from a path \verb+/location/54W/35N/data+,
or logically such as \verb+/data/temperature/snakes+.

This paper contributes a file system abstraction for sensor networks
and a proof-of-concept implementation within ns-2 simulator.
The Plan 9 protocol for implementing the file system abstraction,
Styx, has been already been well-researched on various distributed computing
platforms.
We thus focus our attention on its implementation in sensor networks.

\section{Filesystem Abstraction of Sensor Networks} \label{FSA}

\begin{figure*} 
% [htb]
%\begin{minipage}{.50\textwidth}
% \includegraphics[width=\textwidth]{zoo} 
%\end{minipage}
% \begin{minipage}{.35\textwidth}
\begin{center}
\epsfig{file=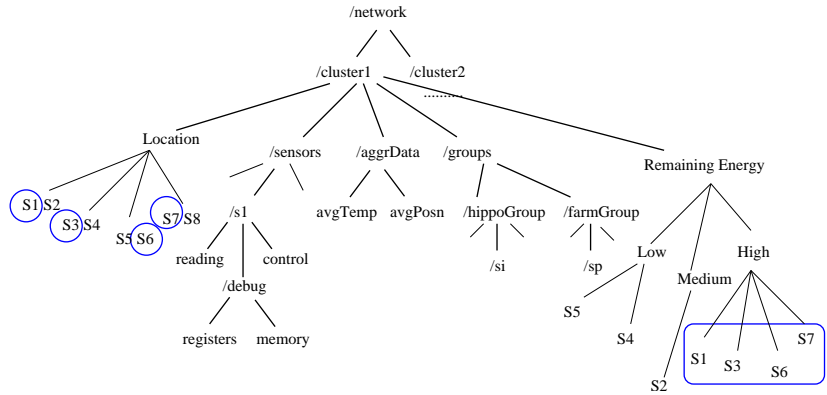,width=0.60\linewidth,silent=}
\end{center}
% \end{minipage}
\caption{Namespace for a sensor network.}
\label{fig:ns}
 \end{figure*}

The central idea of this paper is the application of file system
abstractions for sensor networks. This idea is inspired by the
contributions of the Plan 9 and Inferno operating systems
\cite{pike95plan,dorward1997} whose defining feature was their
consistent treatment of devices and files in a uniform manner. This
section conceptually describes the use of the file system abstraction as
a convenient and scalable means to access, configure and debug
sensor networks.
%% \begin{figure}
%%  \begin{minipage}[t]{8.5cm}
%%  \includegraphics[width=0.9\textwidth]{zoo} 
%%  \end{minipage}
%%  \hfill
%%  \begin{minipage}[t]{8.5cm}
%%  \includegraphics[width=0.9\textwidth]{namespace}
%%  \end{minipage}
%%  \hfill
%%  \end{figure} 
Consider the sample sensor network from a zoo shown in Figure~\ref{fig:ns} with
sensors tracking different animals and resources such as food and
water. The network is divided into two clusters each consisting of a
cluster head. The sensors themselves may each differ in functionality
(temperature/position) and hardware type (MSP/AVR). The associated listing
shows a typical directory layout for such a network.

% \addtolength{\oddsidemargin}{-15mm}  
% \addtolength{\evensidemargin}{-15mm}  
% \addtolength{\oddsidemargin}{+15mm}  
% \addtolength{\evensidemargin}{+15mm} 

The
file representation naturally captures the structure of the network in addition
to depicting logical attributes such as aggregation properties
and groupings. The root directory named \textit{network} encapsulates
the whole network. It has a subdirectory for each of its clusters
which in turn has turn has three subdirectories named
\textit{sensors}, \textit{aggrData}, and \textit{groups}. The \textit{sensors}
directory provides a direct way to access the sensors and has one
directory corresponding to each of the them. Often however, rather
than the individual sensor values, what is of interest is the aggregate
value of a property observed at different sensors. 
The
\textit{aggrData} directory contains files (\textit{avgTemp, avgPosn})
corresponding to these aggregate properties that provide a ready means
to retrieve these cluster wide properties. This is an example of
``intelligence'' being embedded into the file system whereby it is able
to process and interpret data (average from individual readings)
rather than just storing and presenting it. Finally the
\textit{groups} directories demonstrate the file system's ability to
present logical groupings of the sensors according to specific
criteria. The grouping shown is based on animal type, but could
have been based on geographic location of the sensors
(animals). 

The task of locating and naming a sensor device
effectively reduces to finding the path for its corresponding file in
the namespace. Sensor 1's reading for instance, is obtained by reading
\verb+/network/cluster1/sensors/s1/reading+. The low level operations
inherent in retrieving the values are hidden away by the clean file
interface. The representation also easily conceals heterogeneity among
sensors by use of the uniform file interface. Some sensors represented
as part of the network may in fact be simulated while others may be
real. Apart from accessing and reading sensor values, our file
system approach also supports configuration and debugging of sensors. The
file system may provide a \textit{control} file that can
be used to perform control operations on the sensor (e.g., reset,
wakeup, sleep)\ by writing commands to the file. The file system may
also facilitate debugging by exposing the sensors' \textit{registers}
and \textit{memory} as files. An external debugger can then use the
file system interface to debug software executing on the sensors.
% , as
 described in the next section. 

The file system approach offers flexibility in
partitioning application functionality at different levels of the
sensor network (sensor/cluster head/client), which is important
considering that end sensor devices may be computationally
lightweight. Logically combining multiple networks now becomes
analogous to mounting the networks' file system representations under
a common directory.
    
%% \begin{tabular}{|c|} 
%% \hline
%% \begin{figure}[htb]

%% \includegraphics[width=\columnwidth]{zoo}

%% \hspace{.05in}
%% \label{fig:pp}
%% \end{figure}  

%% \end{tabular} \\

%% \begin{figure}[htb]

%% \includegraphics[width=\columnwidth]{namespace}

%% \hspace{.05in}
%% \caption{Namespace for Sensor Net}
%% \label{fig:ns}
%% \end{figure} 

\section{Architecture}
Inspired by the ideas from Plan-9 and Inferno~\cite{pike95plan,dorward1997}, all the
resources are named and accessed like files in a hierarchical file
system and the resources are always accessed via a standard
protocol: Styx. This underlying uniform file system based interface
provides an efficient low-level mechanism on top of which an
application can overlay (possibly an arbitrarily complex) policies
for sensor network representation with the help of namespaces.  In
fact, as shown in the Figure~\ref{fig:ns}, multiple concurrent
perspectives of the network are possible and can co-exist.

An entity that wishes to interact with a sensor network needs to mount
its file system and execute appropriate file operations on it. This
implies that the entity implicitly assumes the role of a file system
client and correspondingly the filesystem implementation assumes
the role of a server. The client and server interact using the Styx
messaging protocol\cite{pike1999} that constitutes encodings of various file
operations. Message exchange always occurs in pairs with the client
initiating the exchange and the server responding. The client starts a
session by connecting to a server using a \textit{Tattach}
message. Once it establishes a connection, the client may navigate the
directory tree using the \textit{Twalk} message (analogous to the
\textit{cd} command). Other standard operations such as opening,
reading and writing to files may be performed using \textit{Topen},
\textit{Tread} and \textit{Twrite} messages respectively. The protocol
supports multiple outstanding requests which is important in the
context of blocking operations. It should be noted that we provide a
software library that client applications may use to talk to the
filesystem. The interface that the library provides consists of
typical file operations (open, read, write, etc.) that the client
application may use to access the file system. Details of the
underlying messaging protocol are completely concealed from the client
application.

The fileserver implementation of sensor networks has
two components in its core, namely device level file servers and
multiplexers. Device level file servers constitute the most basic
forms of servers and are resident in the leaf sensors. They
define a static directory structure and fixed methods for accessing
individual files (really named resources). The files provide the most
basic interactions with the sensors such as reading sensor value and
primitive control operations. Correspondingly, the file servers store
minimal dynamic state about themselves and clients interacting with
them and hence require limited runtime memory.

Multiplexers are
more sophisticated forms of file servers that would typically reside
inside cluster heads of the sensor networks and are responsible for
merging different device level file systems into a cluster level file
system. Multiplexers are stateful servers able to support multiple
client connections and multiple outstanding requests. At startup time
the multiplexer engages in a process of discovery to determine the
topology of the sensors associated with it. It then reads the static
directory structure from the device level file systems of all sensors,
from which the cluster level file system hierarchy is created. When
a client makes a read request on the \textit{reading} file inside one
of the sensor directories, the file server uses the file descriptor in
the \textit{Tread} request to map (multiplex) it to a particular
device file system in its namespace. It then reissues the request to
that device file system to obtain the sensor reading which is then
returned to the client. Supporting outstanding requests means that
even when it is waiting to hear back from the device file system
regarding a request it made, it is still capable of receiving and
processing new requests. 

% \rednote{Nael: I prefer that the aggregate files be maintained by the
% application using mechanisms provided in styx.  This may be intended
% in the description below, but we should make it more explicit.}
A multiplexer has other more involved responsibilities apart from
directing requests to specific device file servers. It is responsible
for supporting aggregate files (\textit{avgTemp \& avgPosn}) in the
cluster file namespace. Since sensor networks often require application specific filters
to be applied on data obtained from various sensors, the file interface uses dynamic libraries to allow the aggregation function used by the multiplexer on the data to be dynamically 
reconfigured as per the application's needs.
The multiplexer is also responsible for managing the
logical grouping of sensors (in \textit{groups} directory) which it
does by appropriately sorting the sensors at the time of enumeration
and at runtime. Migration from using real sensors to a simulation
based setup is straightforward using the multiplexer model. Instead of
being implemented on actual hardware, during simulation the device
file systems get implemented on simulation software while maintaining
the very same file interface.

% \rednote{should we remove the concrete examples section and distribute examples with the description of the filesystem?}

Multiplexers offer great flexibility in
partitioning application, configuration or debugging functionality
among different components of the sensor network. Consider for
instance a debugger client application that is debugging code
executing on a sensor node. It is typical for the debugger to require
access to registers and memory on the sensor in the course of
debugging. Instead of implementing the low level functionality to
retrieve these values inside the debugger itself, the functionality
can instead be implemented in the multiplexer with the cluster file
system providing clients with files for memory and register access. In
this case the debugger can access register/memory indirectly by
reading and writing to these files and have the cluster file system
perform the necessary low level procedures required in performing the
read/write operations.

% \item file server, devices, discovery, virtual directories, drivers, mount device
% \item client library

%  \bibliographystyle{latex8}
%  \bibliography{chiu-1}

% \end{document}
% vim: set sw=2 ts=2 sts=0 expandtab:

  \section{Research Challenges}
In this section, we discuss the research challenges specific to using
the file system abstraction in a sensor network environment.  The
following challenges are identified.

\noindent
{\bf Supporting resource efficient operation:} A key feature of our
  proposed solution is the ability to define namespaces to organize
  the sensor network in an application specific manner. For example,
  as shown in Figure~\ref{fig:ns}, a debugging application can expose the
  sensors' \textit{registers} and \textit{memory} as files. An external
  debugger can then use the file system interface to debug software
  executing on the sensors.  Whereas a data-centric application running
  on the same cluster rather than representing the individual sensor
  values, might represent them in terms of cluster wide properties
  such as aggregate \textit{aggrData} directory contains files
  (\textit{avgTemp, avgPosn}).
  
  Abstractions hide low-level complexities of the underlying system
  and provide rich and intuitive interfaces to the end user, sometimes
  even at the cost of performance/efficiency. However, many
  battery-operated sensors have constraints such as limited energy,
  computational power, and storage capacity, and thus protocols must
  be designed to operate efficiently with these limited resources.
  Therefore in this environment, implementing the file system
  abstraction in a resource efficient manner is essential.  To that
  end, we propose construction of a default resource namespace, that
  exposes resource information (e.g., available energy or
  storage space on a given sensor node) to the applications, helping
  them to make better choices at run time.  An example of
  resource-efficient query execution is presented in
  Section~\ref{sec:examples}.  While the filesystem abstraction 
  provides mechanisms for creating and maintaining namespaces, it 
  does not define how they should be organized (separation of policy 
  from mechanism). 
  
  At this point, we would like to point out that the underlying Styx 
  protocol is lightweight, and therefore the file system abstraction 
  can be implemented over real sensors within reasonable overhead. 
  Further details regarding overhead of the Styx protocol are 
  described in Section~\ref{sec:implementation}.

%  In Section \label{sec:implementation}, we describe in detail
% \rednote{The paragraph below seems like it would be redundant here--we
%  would have already described these things before.  The point about
%  policy and mechanism is over after the first sentence (we can keep
% that and tag it to the end of the above paragraph).}
  
\noindent
\textbf{Consistency models:}
By nature, WSNs are dynamic, concurrent systems.
Thus,
clients' view of the namespace and even data may be inconsistent with
respect to the current actual state.
For example,
a client may use the namespace to determine that a particular mobile sensor
is in a specific region,
only to find that it has actually moved out of that region when it actually reads
data from the sensor.
Or,
stale, cached sensor readings may be sent to a client as a result
of transmission interruptions.
Additional inconsistencies can arise from coupling between different files.
For example,
a client may set a sensor range by writing to a control file,
but subsequently read an out-of-range value from a reading file,
due to caching or other delays.

Strong consistency models could be implemented using distributed
locks and other techniques,
but the nature of WSN applications generally suits weak consistency models.
Sensor data is by nature unreliable,
and applications usually do not rely on high-quality, consistent
operation.
Adopting the file system abstraction,
also allows us to apply research in distributed file system consistency
models such as those developed within the Coda file system project~\cite{kistler-92}.

\noindent
{\bf Managing streaming data:} Sensors typically produce stream data
representing their samples over time. 
% Confirm from Geoffrey and Bhanu about styx and streaming data.
Stream data is not directly supported in the Styx framework;
extensions to the file system abstraction to model streaming devices
may be required. We are currently working on extending
Styx protocol so that streaming data can be handled more efficiently.

\noindent
{\bf Supporting in-network application-specific processing:}
Our framework supports in-network aggregation in the following two
ways. First, a user can extend existing the Styx server to incorporate 
the required functionality. Styx Server implementation is fairly
simple and easy to extend. In the second approach, the user implements
the functionality within an independent module and makes them available
as a dynamic library as described in Section~\ref{FSA}. Then the Styx server loads the dynamic
library at run time when it needs to use those modules and unloads them 
when it is done with the necessary processing (to free up the 
memory resource).
The dynamic library would conform to a defined interface for plug-in
modules.

For applications whose functionality is relatively static, e.g., a 
debugging application (which exposes the set of registers
and memory of sensors'), the first design choice is a better option. Also,
very commonly used aggregation functions including average, min, max 
can be implemented within the Styx server, so that users do not
need to implement these routines themselves. However, for applications
that require more sophisticated in-network processing or whose
functionality changes more often, second design choice is a better 
option. 

\noindent 
{\bf Tolerating network unreliability:} Wireless channels are
susceptible to fading and interference.  Furthermore, 
to conserve energy, sensors are often operated with a low duty cycle
turning off their radios for extended periods of time. 
This intermittent connectivity places unique challenges to filesystem
design.  Fairly static data can be cached and reported while the
sensors are not accessible.  Moreover, cluster-heads can maintain
sensor information to be able to answer queries even when the sensors
are asleep.  

% \noindent
% {\bf
%  Lightweight framework (Styx Overhead):}

% \item Data Consistency and intelligent caching:
% Sensors are enabled with a general purpose microprocessor and therefore handling frequent 
% disconnection etc.... 

% \end{itemize}

  % \rednote{Really need a name and an acronym to avoid saying file system
% abstraction over and over}

% \section{Additional Capabilities of the File System Abstraction}
\section{Additional Capabilities}

Using a file system abstraction offers additional advantages for
application developers in a sensor network.  Some of these are
reviewed in this section.

\noindent
{\bf Ease of application development:} The file system interface is well
understood (both semantically and syntactically) by application
developers and system programmers.  This interface can be easily used
by scientists and researchers who are not familiar with the
intricacies and low-level details of sensor network systems.

%Since now the application developers only
%need to focus on their own application without getting into low
%system-level details, it can facilitate the design and development of
%new applications to a great extent.

\noindent {\bf Access control via file permissions}: File systems
incorporate simple but flexible access control mechanisms via file
permissions.  For example, with the help of appropriate permissions
one can allow only the administrative group members to calibrate
sensors (with write permission), and prevent a normal user from
writing to a sensor by giving them read-only access to the concerned
device. %It provides a way wherein a given authority can delegate
%access control of subparts within a network to the concerned
%authorities.  We believe that the proposed architecture is an elegant
%decentralized security and privacy control mechanism for managing a
%large scale sensor network consisting of tens of thousands of sensors.

\noindent {\bf Ease of integration:} There are many existing
tools designed in other contexts that may be adaptable for use in
sensor network environments because the sensor network is abstracted
as a filesystem.  This includes, for example, development and
visualization tools developed for desktops, PDAs, or even distributed
systems.  Applications of interest can then be ported
over the proposed file system abstraction with an effort significantly
lower than having to develop them from scratch.

\noindent
{\bf Portability across sensor architectures and protocols:} The file
system abstraction using the Styx protocol can serve as a bridging
layer for interoperating heterogeneous sensors as well as interactions
with external devices.  In this sense, it plays a role similar to that
played by IP in interconnecting heterogeneous networks.  Once a new
device has support for file system/Styx primitives, it is able to
interoperate with the remainder of the system.
 
%  with very little
%modification since the filesystem interface is one of the most heavily
%used interface in those applications.  This will give users an
%opportunity to use the tools they are familiar with when dealing with
%these emerging technologies.
%\end{itemize}

  %%From outline:
%% Concrete examples: <shift this and next section to before section IV?>
%%   1. Sensor Monitoring and calibration.
%%   2. Data centric application (DB)
%%   3. Distributed debugging

\section{Examples}\label{sec:examples}

%% 1. Sensor Monitoring and calibration

In this section we demonstrate the use of the file system abstraction
with three examples.  The examples represent important sensor network
functionality and highlight the capabilities of the filesystem
framework.

\subsection{Sensor Monitoring and Calibration}

Monitoring the state of the sensors in terms of their resources is an
important capability for sensor networks~\cite{zhao-02}.  Moreover,
sensor calibration is essential for reducing the noise in the sensor
data~\cite{whitehouse-02}.  The file system provides mechanisms to
discover sensors, as well as read and write their state, which allow
the application developers to rapidly and even interactively monitor
and calibrate the sensor network.  For example, the following commands
can be issued by a client to discover the temperature sensors in an
area, read the remaining energy of one of the sensors and then write a
parameter to calibrate another.

% \rednote{Bhanu, or Sameer can you change these to reflect styx commands and
%   the name space you are going to use}
\begin{verbatim}
mount /dev/network /network
ls /network/cluster1/sensors/
cat /network/cluster1/s1/remaining-energy
echo 2.5 > /network/cluster1/s1/control
\end{verbatim}

%% \rednote{Would like to change this from temperature sensing example
%%   since the next example (aggregation) also uses temperature sensing.
%%   What kind of actuators (other than cooling ones) would be needed
%%   here?}

Note that using an application-specific namespace, we can accomplish
S\&R functionality in a similar manner to the example above.
We may look for the sensors that have a temperature higher than a
threshold, look for actuators near them, and then control the
actuators, for example, to initiate a cooling response in areas that
require it.
\subsection{Data-Centric Application}
%The second example illustrates the use of the filesystem abstraction
%to support an data-centric operation representative of operations in a
%Sense-and-Respond (S&R) systems.  In such a scenario, sensor data may
%be collected and processed in-the-network (e.g., the data may be
%aggregated, or an event detected and a response initiated).  The
%example illustrates: (1) Ability to specify operations (e.g., queries,
%monitoring, reconfiguration) in terms of the application, without
%detailed knowledge of the sensor network infrastructure; (2) the
%ability to execute such operations in the network in a resource
%efficient way.
%algorithm that initiates responses 
%for point number 1, talk about the namespaces, and the ability to
%address sensors using their properties by building namespaces.
%The difference between groupings that are static and those that are
%dynamic.  Consistency models, etc...
%point 2 is basically efficient query execution.  This is addressed in
%two ways.  We have the ability to create a namespace for resources.
%By examining this namespace, we can plan the query execution in a
%resource effeicient way.  The second part is allowing the user to
%cause query execution to occur in the network.  
%The first part is pretty clear to me (at least, to the point where it
%can me described at a high level).  For the second part,  I can do
%the description at the dataflow level as we discussed a few days ago,
%but I am not sure how to describe at the low level.

The second example illustrates the use of the filesystem abstraction
to support a data-centric operation representative of operations in an
S\&R system.  Effective S\&R operation requires in-network processing
to localize interactions and reduce the size of the data transmitted
by the sensors~\cite{estrin-99}.  For example, data from multiple
sensors can be aggregated to reduce the overall data size transported
to an observer.  Conversely, the data may be analyzed to detect events
and initiate responses close to the event location, reducing the cost
of data transmission and enhancing response time.

Consider an example where the temperature in a region (region 10) of
the sensing area is to be monitored, and the average temperature
reported to a monitoring station periodically.  We describe the
planning and execution of this task from a centralized server
perspective for simplicity; however, the namespaces may be maintained,
and the task planning carried out, hierarchically and distributedly by
multiple servers within the sensor network.

First, the application namespaces are consulted to find out what
sensors are available to contribute to the task (for example, using
\verb+ls /network/location/region-10/*+).  This results in discovering
that a cluster (e.g., cluster 1 in Figure~\ref{fig:ns}) is within the 
area of interest. The namespace may now
be consulted to find the actual location of the sensors such that an
appropriate set of sensors in terms of coverage is identified
(e.g., $S_{1}$, and $S_{3} \ldots S_{7}$ in Figure~\ref{fig:ns}).  In
addition, we may consult an energy-based namespace where sensors are
categorized in terms of their remaining energy. This allows
the application to avoid selecting sensors with low available energy
(e.g., $S_{4}$ and $S_{5}$) leaving only high remaining energy sensors
who satisfy the coverage requirements ($S_{1}$, $S_{3}$, $S_6$, $S_7$).
%% in
%% Figure~\ref{figure}).  With the help of above two name spaces,
%% an application can easily select those four sensors which are
%% sufficiently apart from each other, and have high remaining energy
%% ($S_{1}$, $S_{3}$, $S_{6}$, $S_{7}$).

Resource namespaces can be maintained to track network-level resources
to allow query planning in more detail---in our case to decide how to
set up the routing and what points in the network will carry out the
aggregation.  These namespaces may include information regarding the
sensor connectivity, the available bandwidth, and resources committed
to other ongoing tasks.  At the end of this step, the task planning is
accomplished, and a suitable set of sensors, the dataflow in the
network, as well as any in-network processing is determined.

% \rednote{Taking a stab at execution -- Bhanu or Sameer, please feel
%   free to correct}
The query is executed as follows.  The source sensors are tasked with
an appropriate reporting rate (which can later be adapted) to their
upstream neighbors as per the determined dataflow path.  Basic sensors
have support for sending and receiving packets, but some sensors
(e.g., cluster heads) support Styx servers and act as multiplexers.
Communication between the sensors forming the dataflow is set up using
Styx.  
%% Styx allows communication end points to be established using a
%% simple interface.  However, the implementation of the packet handling
%% at either end of the communication is left to the application.
Application specific in-network processing can be accomplished by
customizing packet handlers in these multiplexer nodes.  This can be
done dynamically (allowing specialized handlers to be moved to
appropriate places in the network), statically (at compile time, or
within the Styx protocol), or by allowing the application to select
among a menu of predetermined handlers.

\subsection{Heterogenous Response System Architecture}

\begin{figure}
  \centering
\epsfig{file=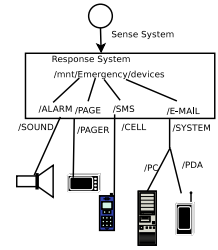,width=0.30\linewidth,silent=}
  \caption{An S\&R system.}
  \label{sandr_diag}
\end{figure}

In the first example, we demonstrated how we can control
actuators embedded with the sensor network to generate
the required response. In this example, we describe
the flexibility of the proposed framework in terms
of incorporating a wide range of heterogenous devices.
As an example, consider a S\&R system (Figure~\ref{sandr_diag}) 
deployed in a chemical factory to detect any gas leakage. 
The response generation system takes input from a range 
of chemical sensors, processes it and then generates the 
necessary response. The response might include local
activities such as controlling actuators embedded within 
the sensor network or it might including contacting 
external entities and authorities or in some cases a 
combination of both.

If the system is built using the file system abstraction, it might have a
directory called /mnt/Emergency and the response generation system
might organize different responses under this directory.  For example,
upon detecting gas leakage, it might set an alarm to alert local
workers and activate the actuators on a sprinkler in order to turn it
off. In addition, it might page or send SMS events to police officers and medical
professionals and e-mail other local authorities.

A task force that manages crisis often consists of individuals from
various government and non-government organizations. In many cases,
the task force is formed in an ad hoc fashion without any knowledge
about underlying sensing infrastructure~\cite{chandy-03}. With the proposed
framework a new device can be mounted on the fly under the
/mnt/Emergency directory and the concerned authority can start getting
the notification messages immediately.  Also, inter-organization
communication can be accomplished more easily using simple navigate,
read, and write commands.  Essentially, the filesystem abstraction
operates as a unifying layer that bridges the differences in the
different underlying organizational networks (much like IP does for
data networks). 

% \rednote{can add a paragraph talking about distributing the
%  namespaces and issues of caching/consistency}.

  \section{Implementation} \label{sec:implementation}
We have a prototype implementation in which we have
integrated the Styx protocol library with the ns-2 simulator~\cite{ns-2}. 
In the current implementation, during the initialization 
phase, the cluster head (CH) discovers the neighboring sensors and
since the CH is running the Styx file server, it simulates 
sensing devices as files in a file system hierarchy. In the current 
implementation, we have incorporated the support for constructing
various namespaces within the Styx server. 
Then the client starts the session with the CH by calling 
the attach function exposed by the client-side Styx library. 
The client-side Styx library then encodes this command into a
low-level Styx message which is sent over the wireless channel. 
The Styx server running on the CH interprets this 
incoming Styx message, processes it, and sends a pointer to 
its root directory to the client, again using the Styx protocol.
It should be noted that, the client-side Styx library 
exposes a clean file system interface and hides all the low-level 
details of the Styx protocol from the client.
Upon getting the pointer to the root directory, the client 
is able to navigate this directory structure using the walk command 
and it reads the files using the read command. In essence, the simulation
set-up supports the capability required by the sensor network
monitoring example described in Section~\ref{sec:examples}. 
In addition, with our simulated prototype, we are able
to simulate a sensor network consisting of at least few hundred sensors.
At present, we are conducting simulations to characterize 
performance of the proposed file system abstraction on a large 
scale sensor network.
 
We also have the basic infrastructure in place for implementing 
the file system abstraction on real sensors such as Berkeley motes. 
To this end, we have developed a lightweight file server model suitable 
for the Motes, which consists of about 1000 lines of code and is less 
than 8KB in size. Our design incorporates the fact that these Motes have 
reasonable about of flash memory (a few KB) but much less RAM (few hundred
bytes), by extensive use of static structures such as device tables and by 
judicious use of dynamic memory. We have also adopted the less demanding 
event-driven model as opposed to using runtime threads. Once this 
implementation is complete we hope to start using it in problems concerning 
resource monitoring, calibration, and distributed debugging all leading to 
more complex data centric applications.

\section{Conclusion}
Sense-and-respond systems are typically heterogeneous
and resource-constrained.
Under these conditions,
system and application development is difficult,
especially for domain experts and other developers whose
specialty may not be embedded systems.
In this paper we have demonstrated how a simple and well-known
abstraction, that of a file system,
hides much of the underlying complexity,
allowing developers to focus on the fundamental challenges
of S\&R systems.
Our initial results with a prototype on the ns-2 simulator
suggest that such an abstraction can be practically implemented.
Our next step is to port our implementation to a physical
WSN such as one constructed from Berkeley Motes.

% LocalWords:  motes

  \bibliographystyle{IEEEtran}
  \bibliography{sensor-net}

\end{document}